\documentclass[letterpaper,preprintnumbers,eqsecnum,superscriptaddress,aps,nofootinbib,11pt]{revtex4}
\usepackage{amssymb}\usepackage[centertags]{amsmath}\usepackage{txfonts}\usepackage{epsfig}\usepackage{bm}
\usepackage{color}\usepackage{graphicx,graphics}\usepackage{multirow}\usepackage{float}\usepackage{ulem}
\usepackage{setspace}\usepackage{slashed}\usepackage{booktabs}\usepackage{hyperref}
\hypersetup{colorlinks=true, citecolor=blue, linkcolor=red, filecolor=black,urlcolor=blue}
\allowdisplaybreaks[4]

\begin{document}

\title{The nuclear matter effect on parton distribution functions}

\author{Weihua Yang}
\affiliation{Department of Nuclear Physics, Yantai University, Yantai, Shandong 264005, China}

\author{Chao Li}
\affiliation{School of Mathematics and Physics, Nanyang Institute of Technology,  Nanyang, 473000, China}


\begin{abstract}

Parton distribution functions (PDFs) are important quantities in describing nucleon structures. They are universal and process-independent. As a matter of fact, nucleon PDFs are inevitably affected by nuclear matter during nuclear scattering process.  In order to study the nuclear PDFs (nPDFs),  in this paper, we introduce the nucleon pair PDFs (dPDFs) to describe parton distributions in the nucleon pair which is confined to a nucleus. We first of all construct the nuclear state in terms of the nucleonic states and calculate the operator definition of nPDFs. Neglecting the higher order corrections or multi-nucleon correlations, we find that nPDFs can be written as a sum of two terms which respectively correspond to PDFs and dPDFs. Nucleon pair PDFs which stem from nucleon-nucleon correlation are proportional to common nucleon PDFs but suppressed by a factor. Compared with the experimental data, we find that dPDFs can explain the behaviour of the EMC effect.  We further calculate the Paschos-Wolfenstein ratio to study the nuclear matter effect on the extraction of weak mixing angle or $\sin^2\theta_W$ by using dPDFs.
\\
\\Keywords: nuclear matter effect, parton distribution functions 

\end{abstract}

\maketitle

\section{Introduction}\label{sec:introduction}

The lepton-nucleon deeply inelastic scattering (DIS) experiment plays an important role in exploring nucleon structures. From the perspective of theoretical calculations,  the cross section of this process can be factorized as a convolution of parton distribution functions (PDFs) and the partonic scattering cross section. Parton distribution functions are universal and essential tools in interpreting experimental data. The one-dimensional PDFs which are used to describe the one-dimensional structures of the nucleon have been known very well due to the large amount of data accumulation \cite{Harland-Lang:2014zoa,Butterworth:2015oua,NNPDF:2017mvq,Hou:2019efy}.  However, the transverse momentum dependent PDFs which are used to describe the three-dimensional structures remain largely unknown. In order to study three-dimensional tomography of the nucleon, future electron-ion colliders are proposed \cite{Accardi:2012qut,AbdulKhalek:2021gbh,Anderle:2021wcy}. More information can be found in Ref. \cite{Boussarie:2023izj}.
As a matter of fact, nucleon PDFs are inevitably affected by nuclear matter in the lepton-nucleus DIS process. In the past three decades, various experiments have confirmed the fact that PDFs measured in free nucleons and nuclei in which nucleons are bounded are significantly different. This implies that nuclei are not simple accumulations of protons and neutrons and they would have non-nucleon degrees of freedom \cite{Frankfurt:1988nt}. Although, different experiments have different kinematic ranges, their data exhibit similar behaviors. For example, the ratio of the cross sections or structure functions between lepton-nuclei and lepton-deuteron scattering processes can be divided into four regions based on the magnitude of the  Bjorken $x$ \cite{Arneodo:1992wf,Guzey:2012yk}. They are respectively known as Fermi motion region ($x>0.8$), EMC effect region ($0.8>x>0.3$) \cite{EuropeanMuon:1983wih,Geesaman:1995yd}, anti-shadowing region($0.3>x>0.1$) and shadowing ($0.1>x$) region. This phenomenon reflects the complexity of the nuclear structure, which can not be explained by a universal theory so far.  

Nuclear PDFs (nPDFs) are introduced to describe nuclei structures. However,  we know even less about them compared to nucleon PDFs \cite{Ethier:2020way}. Nuclear PDFs and PDFs are theoretically assumed to have the same factorization forms and operator definitions except for the replacement of the nucleonic state to the nuclear state. Nuclear PDFs are therefore usually written in the form proportional to nucleon PDFs. Ratios are usually defined as nuclear modification factors. Nuclear matter effects are involved in nuclear modification factors, it is nature to calculate them to determine nPDFs and to further explore nuclear structures in the partonic level. For example, the recent Lattice calculation shows that the PDF exhibits a noticeable difference between the deuteron-like system and the sum of a proton and a neutron for the lattice with a small spatial volume \cite{Chen:2024rgi}. To this end, we present a model to describe nPDFs base on light-cone perturbation theory \cite{Mueller:1989hs,Brodsky:1997de}. The key point is to consider the nucleon-nucleon correlation and introduce the nucleon pair PDFs (dPDFs) to describe parton distributions in the two correlated nucleons or nucleon pair. By considering the operator definitions of these PDFs, we expand the nuclear state under consideration in terms of nucleonic states. Substituting the expansion into the operator definitions of the nPDFs gives the result that nPDFs are expressed as the sum of nucleon PDFs and dPDFs. Nuclear modification factors can  be determined from the dPDFs.
We note that PDFs considered in this paper are known as leading twist unpolarized ones, generally marked as $f_1$ in literatures. Plural forms generally denote different flavours.

Our calculations are based on the fact that two nucleons can form nucleon pair or nucleon-nucleon correlation in a certain nucleus,  which is similar to the short-range correlation (SRC) \cite{Frankfurt:1988nt,Frankfurt:1981mk,Arrington:2011xs,Hen:2016kwk,Fomin:2017ydn,Arrington:2022sov}. For SRC, proton-neutron pairs dominate and are nearly 20 times more than proton-proton pairs \cite{Piasetzky:2006ai,Subedi:2008zz,Hen:2014vua}. Various measurements strongly suggest that there is a linear relation between SRC and the EMC effect \cite{Weinstein:2010rt,Hen:2012fm,Arrington:2012ax,CLAS:2019vsb}. The nucleon-nucleon correlation considered in this paper does not have to be the SRC, but similarities indicate that the EMC effect could be described by using dPDFs in the partonic level. To find applications of dPDFs, we first calculate the ratio of the EMC effect, $R_{EMC}$, which is defined by the nuclear structure functions and deuteron structure functions. 
Compared with the experimental data, we find that dPDFs can describe the behaviour of the EMC effect, but with a deviation of about $4\%$. Uncertainties depend strongly on the nPDFs which are used to determine parameters. 
Furthermore, we calculate the Paschos-Wolfenstein ratio \cite{Paschos:1972kj}, which can be used to extract the weak mixing angle or $\sin^2\theta_W$. We find that the nuclear matter effects are involved in the modification term, $\varepsilon W$. Numerical estimates show that $\varepsilon W$ has a significant impact on the extraction of weak mixing angle.

To be explicit, we organize this paper as follows. In Sec. \ref{sec:fockr}, we introduce the construction of the nuclear state in terms of  the nucleonic states and calculate the operator definition of nPDFs. At the end of this section, we write nPDFs as the sum of nucleon PDFs and dPDFs. In Sec. \ref{sec:slope}, we present applications of the dPDFs by calculating the ratio of EMC effect and the Paschos-Wolfenstein ratio.  We also present numerical estimates of $R_{EMC}$ and the modification term $\varepsilon W$. A brief summary is given in Sec. \ref{sec:summary}.

\section{Representation of the nuclear state} \label{sec:fockr}

The formalism used in this section is based on light-cone perturbation theory  \cite{Mueller:1989hs,Brodsky:1997de}. We would like to expand the physical nuclear state under consideration in terms of the nucleon states of its constituents \cite{Osborne:2002st}. The expansion is different from that in Ref. \cite{Frankfurt:1988nt} where non-nucleonic components, e.g. pion meson, were taken into account.
We first of all introduce the completeness relation, 
\begin{align}
  1= \int \prod_{i=1}^A\frac{dp_i^+d^2\vec p_{i\perp}}{(2\pi)^3 2p_i^+} |\{p_i\} \rangle \langle \{p_i\}| 2p_A^+(2\pi)^3\delta\left(p_A^+-\sum_{i=1}^{A}p_i^+ \right)\delta^2\left(\vec p_{A\perp}-\sum_{i=1}^{A}\vec p_{i\perp}\right), \label{f:completeness}
\end{align}
where momenta are introduced in the light-cone coordinate system. The state $|A\rangle$ is therefore can be rewritten as
\begin{align}
  |A\rangle =\int \prod_{i=1}^A \frac{dp_i^+d^2\vec p_{i\perp}}{(2\pi)^3 2p_i^+}  |\{p_i\} \rangle \phi(\{p_i\}) 2p_A^+(2\pi)^3\delta\left(p_A^+-\sum_{i=1}^{A}p_i^+ \right)\delta^2\left(\vec p_{A\perp}-\sum_{i=1}^{A}\vec p_{i\perp}\right),   \label{f:Aphi}
\end{align}
where $p_i$ is interpreted as the momentum of the $i$th state, and $\phi(\{p_i\})= \langle \{ p_i\} |A\rangle $ is called the light-cone wave function. 
Every state is the eigenstate of $p_A^+, \vec p_{A\perp}$ which satisfies
\begin{align}
 & p_A^+= p^+_i, &&\vec p_{A \perp}=\sum_{i=1}^{A} \vec p_{i\perp}.
\end{align}
 In the nucleus own frame $\vec p_{A
\perp}=\vec 0$. Since $p^+_i>0, p_A^+>0$, we can introduce the fraction $\alpha_i=p_i^+/p_A^+$. The $i$th state momentum can therefore be written as
\begin{align}
  p_i^\mu=\left(\alpha_ip_A^+, \frac{\vec p^2_{i\perp}+m_i^2}{2\alpha_ip_A^+}, \vec p_{i\perp}\right),
\end{align}
where $m_i$ is the mass of the $i$th state. We note that the minus component of each momentum is determined by the on-shell condition, it does not have to be conserved. The state is normalized to the delta function:
\begin{align}
  \langle \{p_i\}| \{p'_i\}\rangle =\prod_{i=1}^A2p_i^+ (2\pi)^3 \delta^3(p_i-p'_i).
\end{align}

A nucleus can be seen as a bound state of nucleons which are bounded together by nuclear force.  For the leading order approximation, nucleons can be seen as $free$ particles in the nucleus. This allows us to only consider the lowest order nucleonic state. In other words, $|\{p_i\}\rangle$ in Eq. (\ref{f:Aphi}) are free nucleonic states.  For the next-to-leading order corrections, we consider the nucleon-nucleon correlations and $|\{p_i\}\rangle$ would include nucleon pair state. Higher order corrections are neglected. 


\subsection{Free state representation}

Parton distribution functions are defined through two points correlation functions or quark-quark correlation functions. To obtain the operator definition of the nPDF, we present the two points correlation function defined through the nuclear state vector $|A\rangle$,
\begin{align}
  \Phi(k)=\int d^4\xi e^{ik\xi} \langle A |\bar\psi(0) \psi(\xi)|A\rangle.     \label{f:nphi}
\end{align}
For the sake of simplicity, the gauge link that has no influence on calculations has been neglected. The nPDF can be calculate via the following formula,
\begin{align}
  2f^A(x)p_A^\mu = \int\frac{d^4k}{(2\pi)^4}\delta\left(x-\frac{k^+}{p_A^+}\right)\rm{Tr}\left[\gamma^\mu \Phi(k)\right].
\end{align}
Superscript $A$ denotes nucleus, $k$ is the momentum of the incident parton and $p_A$ is the momentum of the nucleus. Without loss of generality, we can set $\mu=+$ to obtain
\begin{align}
  f^A(x)= \frac{1}{2\pi}\int d\xi e^{ixp_A^+\xi^-} \langle A |\bar\psi(0)\frac{\gamma^+}{2} \psi(\xi^-)|A\rangle.    \label{f:npdf}
\end{align}

To find out relations between nucleon PDFs and nPDFs, we substitute Eq. (\ref{f:Aphi}) into (\ref{f:npdf}) and only consider the free nucleonic state components. After  a simple mathematical calculation, we have
\begin{align}
  f^A(x)& = \frac{1}{2\pi}\int d\xi e^{ixp_A^+\xi^-} \int DP'\int DP\phi^*(p'_i)  \phi(p_i) \langle p'_i|\bar\psi(0)\frac{\gamma^+}{2} \psi(\xi^-) |p_i\rangle  \nonumber \\
  &=A \int \frac{d\alpha}{\alpha} \hat\rho(\alpha) f^N(x/\alpha), \label{f:fafree}
\end{align}
where coefficient $A$ is the mass number. Momentum fraction $\alpha$ is defined by $\alpha=p^+/p_A^+$, $p$ is the momentum of the certain nucleon to which the incident parton belongs. 
In Eq. (\ref{f:fafree}),
\begin{align}
  \int DP\equiv\int \prod_{i=1}^{A}\frac{dp_i^+d^2\vec p_{i\perp}}{(2\pi)^2 2p_i^+} 2p_A^+(2\pi)^3\delta\left(p_A^+-\sum_{i=1}^{A}p_i^+\right)\delta^2\left(\vec p_{A\perp}-\sum_{i=1}^{A}\vec p_{i\perp}\right)
\end{align}
and  $\hat\rho(\alpha)$ is defined as 
\begin{align}
   \hat\rho(\alpha)&=\frac{1}{4\pi\alpha} \int\frac{d^2\vec p_\perp}{(2\pi)^2} \rho_1(p), \label{f:rhoalpha}\\
   \rho_1(p)&=\int \prod_{j=1}^{A-1}\frac{dp_j^+d^2\vec p_{j\perp}}{(2\pi)^3 2p_j^+}| \phi(p,\{p_j\}) |^2 2p_A^+(2\pi)^3  \delta\left(p_A^+ - p^+-\sum_{j=1}^{A-1}p_j^+\right)\delta^2\left(\vec p_\perp +\sum_{j=1}^{A-1}\vec p_{j\perp}\right). \label{f:rhop}
\end{align}
 $\hat \rho(\alpha)$ can be seen as the longitudinal distribution function or the probability of finding a nucleon with momentum fraction $\alpha$ in the nucleus. $\rho_1(p)$ is the distribution function or the probability of finding a nucleon with momentum $p$ in the nucleus. It satisfies \cite{Osborne:2002st}
\begin{align}
  \int\frac{d^3\vec p}{(2\pi)^32p^+} \rho_1(p)=1.
\end{align}
The operator definition of the distribution function $ f^N(x/\alpha)$ is
\begin{align}
  f^N(x/\alpha) &= \frac{1}{2\pi}\int d\xi e^{ixp_A^+\xi^-}  \langle p|\bar\psi(0)\frac{\gamma^+}{2} \psi(\xi^-) |p\rangle \nonumber\\
  &=\frac{1}{2\pi}\int d\xi e^{ixp^+\xi^-/\alpha}  \langle p|\bar\psi(0)\frac{\gamma^+}{2} \psi(\xi^-) |p\rangle.
\end{align}
Superscript $N$ denotes nucleon. We notice from Eq. (\ref{f:fafree}) that the unpolarized nPDF $f^A(x)$ is proportional to the mass number $A$ and also to the integration of the distribution function $f^N(x/\alpha)$ multiplying the nucleon distribution function $\hat{\rho}(\alpha)$.  In other words, nuclei are simple accumulations of protons and neutrons at the leading order and $f^A(x)$ is just the sum of the $A$ distribution functions $f^N(x/\alpha)$ weighted by distribution functions $\hat{\rho}(\alpha)$.

We also note here that it is possible to choose two nucleons at the same time and there are $C^2_A=A(A-1)/2$ combination forms. Similarly, one can define $\rho_2(p_1,p_2)$, just like $\rho(p)$ in Eq. (\ref{f:rhop}), to represent the probability of finding two nucleons with the specified momenta within the nucleus. However, we do not consider this case in this paper because it is a result of the double nucleon scattering process \cite{Osborne:2002st}.

\subsection{Nucleon pair state representation}

When considering the nucleon pair state, ignoring the detailed procedure for simplicity, $f^A(x)$ can be written as 
\begin{align}
  f^A(x)& =n^A \int \frac{d\beta}{\beta} \hat\rho(\beta) f^D(x/\beta), \label{f:fapair}
\end{align}
where $n^ A$ is the number of nucleon pair and $\beta=D^+/p_A^+$ with $D$ the momentum of the nucleon pair. Superscript $D$ denotes nucleon pair. The relevant nucleon pair distribution functions are given by
\begin{align}
   \hat\rho(\beta)&=\frac{1}{4\pi\beta} \int\frac{d^2\vec D_\perp}{(2\pi)^2} \rho_2(D), \\
   \rho_2(D)&=\int \prod_{j=1}^{A-2}\frac{dp_j^+d^2\vec p_{j\perp}}{(2\pi)^3 2p_j^+}| \phi(D,\{p_j\}) |^2 2p_A^+(2\pi)^3  \delta\left(p_A^+ - D^+-\sum_{j=1}^{A-2}p_j^+\right)\delta^2\left(\vec D_\perp +\sum_{j=1}^{A-2}\vec p_{j\perp}\right),
\end{align}
where $\rho_2(D)$ satisfies
\begin{align}
\int \prod_{i=1}^{2}\frac{d^3\vec p_i}{(2\pi)^32p_i^+} \rho_2(D)=1.
\end{align}
The operator definition of the distribution function $ f^D(x/\beta)$ is given by
\begin{align}
  f^D(x/\beta) &= \frac{1}{2\pi}\int d\xi e^{ixp_A^+\xi^-}  \langle D|\bar\psi(0)\frac{\gamma^+}{2} \psi(\xi^-) |D\rangle \nonumber\\
  &=\frac{1}{2\pi}\int d\xi e^{ixD^+\xi^-/\beta}  \langle D|\bar\psi(0)\frac{\gamma^+}{2} \psi(\xi^-) |D\rangle. \label{f:fDx}
\end{align}
The number of nucleon pair $n^A$ shown in Eq. (\ref{f:fapair}) do not have to be equal to $C^2_A= A(A-1)/2$ because not all nucleons would form nucleon pairs due to the nuclear force. This is the reason we insert $n^A$ rather than $A(A-1)/2$ in Eq. (\ref{f:fapair}).


To find relations between the distribution function $f^D$ in the nucleon pair and distribution function $f^N$ in the nucleon, we present an intuitive method  based on the arguments given below.
\begin{itemize}
  \item The state of the nucleon pair is described by the state vector $|D \rangle $ which can be written as a function of the distance $r$, i.e. $|D\rangle = |D(r) \rangle$. $r$ is the radial distance between the two nucleons in the pair.
  \item Two nucleons correlate to each other if $r<R$, $R$ being the critical distance of the correlation. We do not know the exact value, but it is reasonable to be twice the radius of a nucleon, $R\sim 1.7$ fm. If $r\geq R$, the correlation vanishes and  $|D(r) \rangle $ is reduced to free particle states. $|D(R) \rangle = |p_1 \rangle|p_2 \rangle$. It is convenient to normalize the free particle states to unit, $\langle p_i|p_j \rangle =\delta_{ij}, i,j = 1,2$.
\end{itemize}
These two arguments can be alternatively given in the momentum picture and they will leading to the similar results obtained in the following. We do not show them here for simplicity.
According to the two arguments, we make an expansion around $R$ for $f^D(x)$ and obtain
\begin{align}
  f^D(x_D)= &f^D(x_D)|_R+(r-R)\frac{d}{dr}f^D(x_D)|_R   + \frac{1}{2!} (r-R)^2\frac{d^2}{dr^2}f^D(x_D)|_R +\cdots \nonumber \\
  = & e^{i\hat{y}\hat{p}_r}f^D(x_D)|_R, \label{f:tailor}
\end{align}
where $\hat{y}=R-r, \hat p_r =i \frac{d}{dr}$ are respectively the position and momentum operators. $\hat{p}_r$ only acts on $|D\rangle$.  Subscript $D$ is used here to distinguish fraction $x_D$ from $x$. We note that it is more convenient to reduce $\hat y$ to a one-dimensional case. For the given operator definition given in Eq. (\ref{f:fDx}),
we obtain
\begin{align}
   f^D(x_D)=2\cos\left(yp_r\right) f^N(x_D/\gamma), \label{f:fxemc}
\end{align}
where $\gamma=p^+/D^+$, $p$ is the momentum of the chosen nucleon in the pair. $p_r$ is the difference of the two nucleon momenta. We notice that 
\begin{align}
  f^D(x_D)=2f^N(x_D/\gamma) \label{f:fD2}
\end{align}
when $r=R$. However, if $r<R$, $f^D(x_D)$ is inevitable suppressed by a factor.



From previous discussions, we notice that the complete $f^A(x)$ includes two parts, $f^N$ and $f^D$, if higher order corrections are neglected. The complete unpolarized nPDF is therefore a sum of these two terms,
\begin{align}
  f^A(x)&=(A-2n^A) \int \frac{d\alpha}{\alpha} \hat\rho(\alpha) f^N(x/\alpha) + n^A \int \frac{d\beta}{\beta} \hat\rho(\beta) f^D(x/\beta). \label{f:fasum}
\end{align}
Nucleons in the nucleon pairs are not free any more. Therefore the number of free nucleons should be $(A-2n^A)$ rather than $A$. The factor $2$ before $n^A$ is due to the existence of two nucleons in the pair. 
One would not use the same method to calculate $f^D$ otherwise Eq. (\ref{f:fasum}) would reduce to Eq. (\ref{f:fafree}). 

From  Eq. (\ref{f:fasum}) we notice that $f^A(x)$ is a function of fraction $x$, the ratio of the incident quark momentum to the nucleus momentum. It is not a direct addition of the nucleon parton distribution ($f^N$)  and the nucleon pair parton distribution ($f^D$) due to nuclear matter effects or mathematical integrations. Nuclear effects are just hidden in nucleon and nucleon pair distributions, $\hat \rho(\alpha)$ and $\hat\rho(\beta)$.
Let us reconsider Eq. (\ref{f:fasum}) and take the first term for example. $\hat{\rho}(\alpha)$ is known as the longitudinal distribution function of the chosen nucleon with momentum fraction $\alpha$ in a nucleus and $f^N(x/\alpha)$ is distribution function of the certain parton in the nucleon. From another perspective, we can take $\hat{\rho}(\alpha)/\alpha$ as the distribution function of $f^N(x/\alpha)$ in the $\alpha$ space. In other words, $\hat{\rho}(\alpha)/\alpha$ describes the probability of finding $f^N(x/\alpha)$ in the $\alpha$ space. Integrating over $\alpha$ in the $\alpha$ space therefore gives the $real$   PDF regardless of the nucleus. This is also the reason why we use `distribution  function' rather than `PDF' in the previous context.  This idea builds a bridge between PDF and nPDF. The same argument applies to $ f^D(x/\beta)$. In this case, we can rewrite $f^A(x)$ as
\begin{align}
  f^A(x)&=(A-2n^A)  f^N(x)+ n^A  f^D(x), \label{f:fareal}
\end{align}
where $f^N(x)$ and $ f^D(x)$ are defined as the PDF(s) and dPDF(s), respectively,
\begin{align}
  &  f^N(x)\equiv \int \frac{d\alpha}{\alpha} \hat\rho(\alpha) f^N(x/\alpha) , \\
  &  f^D(x)\equiv \int \frac{d\beta}{\beta} \hat\rho(\beta) f^D(x/\beta).
\end{align}

To obtain the explicit relationship between $f^A(x)$ and $ f^N(x)$, we assume that the nPDF $f^D(x)$ is proportional to $ f^N(x)$ and can be written as 
\begin{align}
  f^D(x)=2 (1-\Lambda) f^N(x),  \label{f:fddelta}
\end{align}
where parameter $\Lambda$ is introduced for convenience. We therefore finally have
\begin{align}
  f^A(x)&=(A-2n^A) f^N(x)+2 n^A  (1-\Lambda) f^N(x)\nonumber\\
  &=(A-2n^A \Lambda)f^N(x). \label{f:farealfinal}
\end{align}
From this equation, we notice that $ f^A(x)$ can be determined by $ f^N(x)$ as long as parameters $\Lambda$ and $n^A$ are known. 
Here we note $f^A(x)$ is usually written as $Af^A(x)$ in literatures to consistent with structure functions. Similar conventions will be given in the following section. 

\section{Applications} \label{sec:slope}

Equation (\ref{f:fareal}) provides a simple and clear interpretation of nuclear matter effects by including contributions from dPDFs. These effects, according to our calculation, are involved in factors $\Lambda$ and $n^A$. In this section, we present applications of the dPDFs by calculating the EMC effect \cite{EuropeanMuon:1983wih} and the Paschos-Wolfenstein ratio \cite{Paschos:1972kj}.

\subsection{The EMC effect}

The EMC effect was observed four decades ago and is believed to be due to the modification of the valance quarks in the nuclear medium. It states that the electron DIS cross sections of nuclei ($A\geq 3$) are smaller than those of deuterium at moderate $x$, $0.3<x<0.8$. 
In this part, we will use the dPDFs to calculate the EMC effect.  Different from previous modifications of structure function $F_2$, see formulae in Ref. \cite{CLAS:2019vsb}, we here present calculations at partonic level.

The EMC effect is defined by the ratio of the nuclear structure function to the deuteron structure function,
\begin{align}
  R_{EMC}=\frac{F_2^A}{F_2^d}. \label{f:Remcdef}
\end{align}
We define the nuclear structure function $F_2^A$ via the following expression,
\begin{align}
  A F_2^A = Z F_2^{p/A} + N F_2^{n/A},  \label{f:AF2A}
\end{align}
where superscripts $(p,n)/A$ denote the proton and neutron in the given nucleus. We assume that $F_2^A$ has the same form to $F_2^{p,n/A}$ when they are written in terms of corresponding quark distribution functions, i.e.
\begin{align}
 & F_2^A = x\left(Q_u^2 u^A + Q_d^2 d^A\right), \label{f:F2A}\\
 & F_2^{p/A} = x\left(Q_u^2 u^{p/A} + Q_d^2 d^{p/A}\right), \\
 & F_2^{n/A} = x\left(Q_u^2 u^{n/A} + Q_d^2 d^{n/A}\right). \label{f:F2nA}
\end{align}
Here $Q_u$ and $Q_d$ are charge numbers of the up and down quarks. $u^A, d^A$ are the up and down quark distribution functions in the nucleus  while $u^{p/A} (u^{n/A})$ and $d^{p/A} (d^{n/A})$ are the up and down quark distribution functions in a proton (neutron) bounded inside the given nucleus. Only valence quarks are taken into account in our calculations.
From Eqs. (\ref{f:AF2A})-(\ref{f:F2nA}), we have
\begin{align}
& Au^A =Z  u^{p/A}  + N u^{d/A}, \label{f:uAA}\\
& Ad^A =Z  d^{n/A}  + N d^{n/A}. \label{f:dAA}
\end{align}

Experimental data have shown that $u^{p,n/A}$ or $d^{p,n/A}$ are different from $u^{p,n}$ or $d^{p,n}$ in free nucleons. According to calculations in the previous section,  nuclear parton distributions can be given in terms of nucleonic parton distributions, 
\begin{align}
  & Zu^{p/A} =(Z-n^A) u^p+ n^A  \tilde u^p, \label{f:ZupA} \\
  & Zd^{p/A} =(Z-n^A) d^p+ n^A  \tilde d^p, \label{f:ZdpA} \\
  & Nu^{n/A} =(N-n^A) u^n+ n^A  \tilde u^n, \label{f:NunA} \\
  & Nd^{n/A} =(N-n^A) d^n+ n^A  \tilde d^n. \label{f:NdnA}
\end{align}
Here $u^{p,n}$ and $d^{p,n}$ are up and down distribution functions in free nucleons. $\tilde u^{p,n}$ and $\tilde d^{p,n}$ are nucleon pair distribution functions. 
From Eq. (\ref{f:fddelta}), they can be rewritten as
\begin{align}
  \tilde{u}^{p,n}&=(1-\Lambda)u^{p,n}, \\
  \tilde{d}^{p,n}&=(1-\Lambda)d^{p,n}.
\end{align}
 Introducing the following definitions,
\begin{align}
  & \delta u^p = \tilde u^p-u^p, \label{f:deltaup}\\
  & \delta u^n = \tilde u^n-u^n, \\
  & \delta d^p = \tilde d^p-d^p, \\
  & \delta d^n = \tilde d^n-d^n,  \label{f:deltadn}
\end{align}
we obtain,
\begin{align}
  & u^{p/A} =\frac{1}{Z}\left(Z u^p+ n^A  \delta u^p \right), \label{f:upA} \\
  & d^{p/A} =\frac{1}{Z}\left(Z d^p+ n^A  \delta d^p \right), \label{f:dpA} \\
  & u^{n/A} =\frac{1}{N}\left(Nu^n+ n^A  \delta u^n \right), \label{f:unA} \\
  & d^{n/A} =\frac{1}{N}\left(N d^n+ n^A  \delta d^n \right). \label{f:dnA}
\end{align}
Substituting these equations and using Eqs. (\ref{f:F2A})-(\ref{f:F2nA}), we have
\begin{align}
  AF_2^A =&Zx\left[Q_u^2\frac{1}{Z}\left(Z u^p+ n^A  \delta u^p \right) +Q_d^2 \frac{1}{Z}\left(Z d^p+ n^A  \delta d^p \right)\right] \nonumber \\
  + & Nx \left[Q_u^2\frac{1}{N}\left(Nu^n+ n^A  \delta u^n \right)+Q_d^2 \frac{1}{N}\left(N d^n+ n^A  \delta d^n \right)\right] \nonumber\\
  = & x Q_u^2 \left(Z u^p +N d^p\right) +x Q_d^2 \left(Z d^p +N u^p\right) \nonumber\\
   +& x Q_u^2 n^A\left(\delta u^p +\frac{N}{Z} \delta d^p\right) +x Q_d^2 n^A\left(\delta d^p +\frac{N}{Z} \delta u^p\right). \label{f:AF2PDF}
\end{align}
Here we have used the isospin symmetry \cite{Yang:2023zmr},
\begin{align}
  & u^p=d^n,   \quad \quad  u^n=d^p, \label{f:isoud} \\
  & \delta d^n=\frac{N}{Z} \delta u^p,  \quad \quad \delta u^n=\frac{N}{Z} \delta d^p. \label{f:isoudd}
\end{align}

With the same way we can write down the structure function of the DIS of the deuterium,
\begin{align}
  2F_2^d & =  x \left(Q_u^2 +Q_d^2\right)\left( u^p + d^p\right) + x  n^d\left(Q_u^2 +Q_d^2\right)\left(\delta u^p +\delta d^p\right), \label{f:AF2D}
\end{align}
where $ n^d$ is the number of nucleon pairs in a deuteron.

Substituting Eqs. (\ref{f:AF2PDF}) and (\ref{f:AF2D}) into Eq. (\ref{f:Remcdef}), we obtain
\begin{align}
  R_{EMC}=\frac{2}{A}\frac{1-\frac{n^A}{Z}\Lambda }{1-n^d\Lambda }\frac{X(u,d)}{Y(u,d)}, \label{f:Remc}
\end{align}
where
\begin{align}
  &X(u,p)=Q_u^2\left(Z u^p+ Nd^p \right) +Q_d^2\left(Z d^p+ N u^p \right), \label{f:Xud} \\
  &Y(u,p)=\left(Q_u^2+Q_d^2\right)\left(u^p+ d^p \right). \label{f:Yud}
\end{align}
With the ratio $R_{u,d}^A=f_i^{p/A}/f_{u,d}$, one can calculate $\Lambda$ by using Eqs. (\ref{f:upA})-(\ref{f:dnA}), 
\begin{align}
  \Lambda=\frac{Z}{n^A} (1-R_{u,d}^A).  \label{f:Rud}
\end{align}

\begin{figure}
  \centering
  \includegraphics[width=0.5\linewidth]{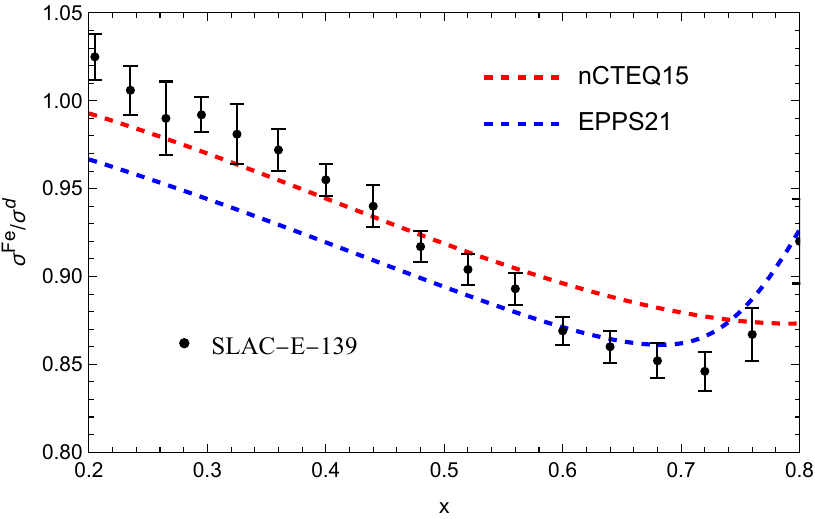}
  \caption{Estimations of $R_{EMC}$ with respect to momentum fraction $x$ at 6 GeV. For the red dashed line, $n^d Z/n^A$ is taken as 0.88. For the blue dashed line, $n^d Z/n^A$ is taken as 0.91.}\label{fig.sigma}
\end{figure}

To test Eq. (\ref{f:Remc}), we present numerical estimates in Fig. \ref{fig.sigma}, where the ratio of the cross section of iron and that of deuterium is shown \cite{Gomez:1993ri}. The red dashed line shows result based on the nCTEQ15  \cite{Kovarik:2015cma}, which are used to extract parameters by using Eq. (\ref{f:Rud}). We note that  $Z/n^A$  is not known and we parameterize $Z/n^A$ and $n^d$ together,  $n^d Z/n^A=0.88$.  The blue dashed line shows result based on the EPPS21 \cite{Hou:2019efy,Eskola:2021nhw} and  $n^d Z/n^A=0.91$. From Fig. \ref{fig.sigma}, we notice Eq. (\ref{f:Remc}) can describe the behaviour of the EMC effect, but there are still some deviations.  On the one hand, parameters extracted from nPDFs have larger uncertainties. More data is needed to improve the accuracy of nPDFs. On the other hand, the model used in this paper may not describe all details of the EMC effect. Because $\Lambda$ is taken as a constant in our consideration. We could study the $x$-dependence of $\Lambda$ or introduce one more parameter in Eq. (\ref{f:fddelta}) to improve this model for future studies.


\subsection{The Paschos-Wolfenstein ratio}

The Paschos-Wolfenstein ratio is defined via the neutral current and charged current neutrino and antineutrino DIS of nucleon. The expression is given by \cite{Paschos:1972kj}
\begin{align}
R^{-}=\frac{\sigma_{N C}^{\nu N}-\sigma_{N C}^{\bar{\nu} N}}{\sigma_{C C}^{\nu N}-\sigma_{C C}^{\bar{\nu} N}}. \label{f:pwn}
\end{align}
The result in standard model is $\frac{1}{2}-\sin^2\theta_W$. Hence, it can be used to extract the weak mixing angle or $\sin^2\theta_W$. However, $R^-$ would be different in the DIS process of nuclei where $N\neq Z$ \cite{Cloet:2009qs}. In this part, we recalculate this ratio  with dPDFs to study the nuclear matter effect on the extraction of the weak mixing angle.

For the sake of simplicity, we do not show the related cross sections which can be found in Ref. \cite{Kumano:2002ra}, but only present the definition of the Paschos-Wolfenstein ratio in terms of nPDFs,
\begin{align}
  R^-=y(2-y)\frac{w_1 u^A+w_2 d^A}{d^A-(1-y)^2u^A}, \label{f:pwnuclei}
\end{align}
where $y$ is the kinematic variable in DIS process and defined as $y=(p\cdot q)/(p\cdot l)$, see Ref. \cite{Yang:2022sbz} for example. $w_1, w_2$ are introduced as
\begin{align}
  & w_1=\frac{1}{4}-\frac{2}{3}\sin^2\theta_W, \\
  & w_2=\frac{1}{4}-\frac{1}{3}\sin^2\theta_W.
\end{align}
They satisfy $w_1+w_2=\frac{1}{2}-\sin^2\theta_W$. Using Eqs. (\ref{f:ZupA})-(\ref{f:NdnA}) and (\ref{f:upA})-(\ref{f:dnA}), one obtains
 \begin{align}
  R^-=\frac{1}{2}-\sin^2\theta_W+\varepsilon W, \label{f:pwfinal}
\end{align}
where $\varepsilon=(N-Z)/Z$ and
\begin{align}
  W=\frac{y(2-y)\left(w_1 d^p +w_2 u^p\right)-\left[u^p-(1-y)^2d^p\right](w_1+w_2)}{y(2-y)\left( d^p +u^p\right)+\varepsilon\left[u^p-(1-y)^2d^p\right]}.
\end{align}
In order to obtain Eq. (\ref{f:pwfinal}), we have used the isospin symmetry given in Eqs. (\ref{f:isoud}) and (\ref{f:isoudd}) and neglected contributions from strange quarks and charm quarks.

\begin{figure}[t]
  \centering
  \includegraphics[width=0.5\linewidth]{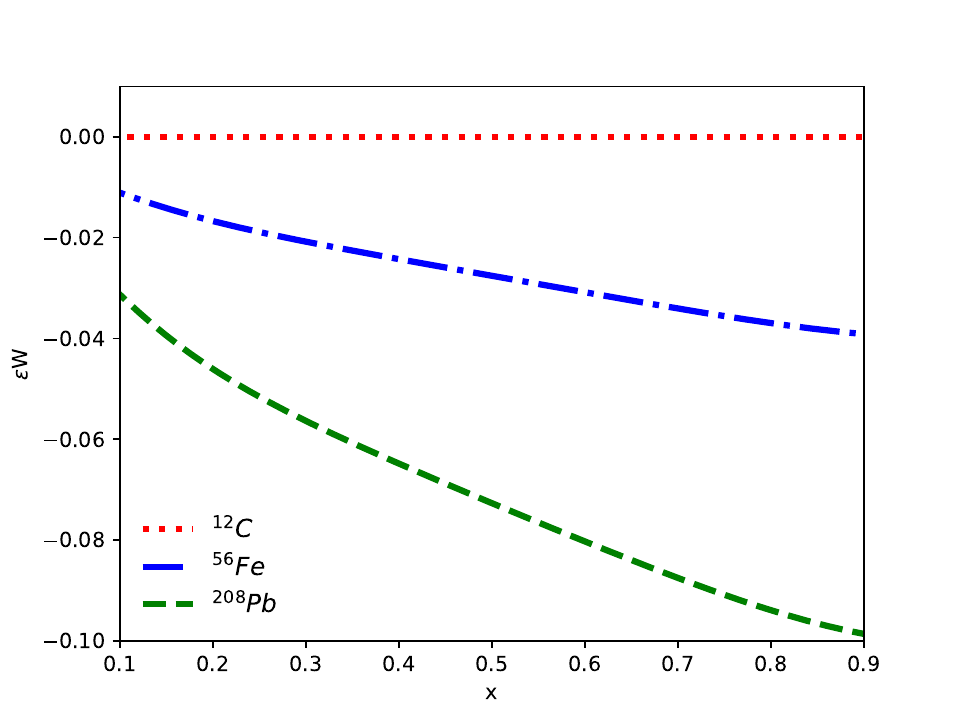}
  \caption{Estimations of $\varepsilon W$ with respect to momentum fraction $x$ at $5$ GeV.  The red, blue and green lines respectively correspond to $^{12}C,  ^{56}Fe $ and  $^{208}Pb$.}\label{fig:ewx}
\end{figure}

\begin{figure}[t]
  \centering
  \includegraphics[width=0.5\linewidth]{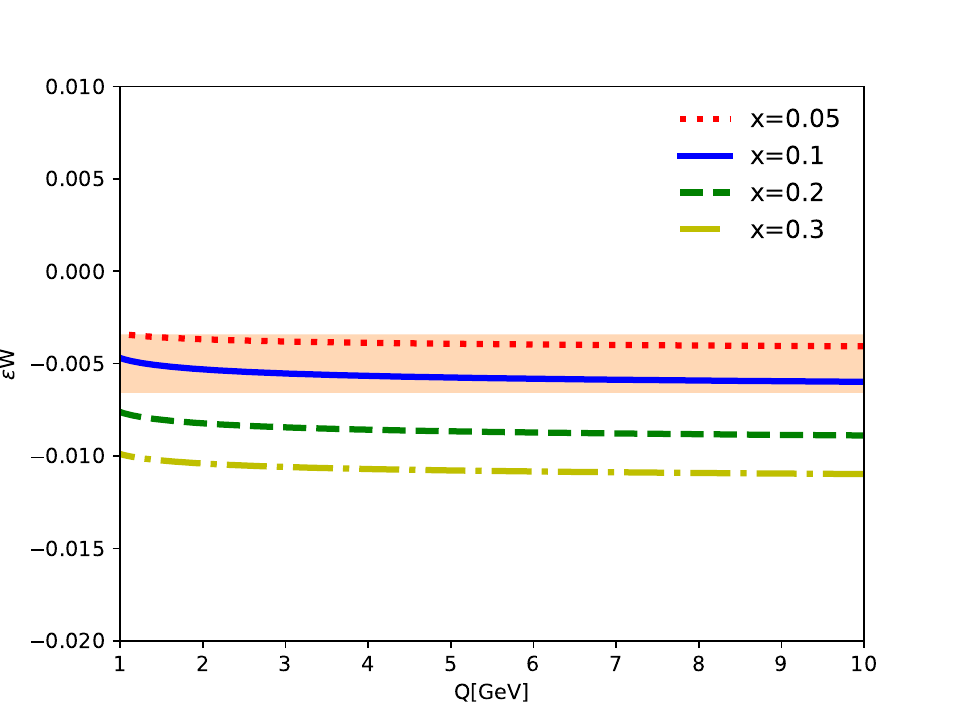}
  \caption{Estimations of $\varepsilon W$ for iron with respect to $Q$ at different $x$. The central value $-0.005$ is the difference of $\sin^2\theta_W$ between the expected value and the NuTeV measurement \cite{NuTeV:2001whx}. The brown band denotes the error band. }\label{fig:ewQ}
\end{figure}

From Eq. (\ref{f:pwfinal}) we can see that nuclear matter effects on Paschos-Wolfenstein ratio are involved in $\varepsilon W$. $\varepsilon$ indicates the neutron excess effect that nuclear modifications vanish in the symmetric nuclei \cite{Cloet:2009qs} while $W$ includes corrections from quark distribution functions. In Figs. \ref{fig:ewx} and \ref{fig:ewQ} we show numerical estimates of the modification factor $\varepsilon W$ with respect to the $x$ and $Q$, respectively. The quark distribution functions are taken from CT14 \cite{Dulat:2015mca}. The kinematic variable $y$ is set as $0.5$. In Fig. \ref{fig:ewx}, we see that the nuclear matter effects are more significant in nuclei with more neutrons. For further understanding the behaviour of the modification factor, we show the numerical estimates of iron in Fig. \ref{fig:ewQ}. The central value $-0.005$ is the difference of $\sin^2\theta_W$ between the expected value and the NuTeV measurement \cite{NuTeV:2001whx}. The brown band denotes the error band.

According to these numerical estimates, we notice that nuclear matter effect do have a significant impact on the extraction of the weak mixing angle in the framework of dPDFs. In Fig. \ref{fig:ewQ}, numerical results show that dPDFs can be used to explain the discrepancy between the NuTeV measurement and the expected value, at least partly. Our results are similar to that in Ref. \cite{Yang:2023zmr} but do not depend on the parametrizations of nPDFs (In Ref. \cite{Yang:2023zmr}, $\Delta^-$ term was given averaged on $x$). Because the modification factor $\varepsilon W$ only includes $u^p$ and $d^p$ and excludes nPDFs.



\section{Summary}\label{sec:summary}

Parton distribution functions are important quantities in describing nucleon structures. However, nucleon PDFs are inevitably affected by nuclear matter in the lepton-nucleus DIS process. Various experiments have also confirmed the fact that PDFs measured in free nucleons and in the nuclei in which nucleons are bounded are significantly different.  In order to describe nPDFs or to study the nuclear matter effects, we introduce the dPDFs by using light-cone perturbation theory in this paper. According to our calculation, we find that the nPDFs can be written as the sum of the PDFs and dPDFs and dPDFs are proportional to PDFs. By utilizing dPDFs, we calculate the EMC effect and notice that  Eq. (\ref{f:Remc}) can describe the behaviour of the EMC effect, but there are still some deviations.  Furthermore, we recalculate the Paschos-Wolfenstein ratio with dPDFs to study the nuclear matter effect on the extraction of weak mixing angle. The numerical estimates show that nuclear matter effect do have a significant impact on that in the framework of dPDFs. We also find that dPDFs can explain the discrepancy between the NuTeV measurement and the expected value. 

The introduction of dPDFs is based on the fact that nucleons can form nucleon-nucleon correlations or nucleon pairs due to nuclear force. However, several questions which were not explained in this paper need to be further considered in future studies. For example, the relation between the SRC and the nucleon-nucleon correlation should be illustrated. SRC describes nucleon pairs in nuclei with high relative momentum due to tensor force. What is the dynamics of the nucleon-nucleon correlation given in this paper? In addition, is it reasonable to introduce dPDFs?  The recent lattice calculation seems to show a positive answer \cite{Chen:2024rgi}. However, more researches are needed.

\section*{ACKNOWLEDGEMENTS}
The authors thank Ji Xu and Xinghua Yang very much for helpful discussions and kindly hospitalities. This work was supported by the National Natural Science Foundation of China (Grant No. 12405103), Natural Science Foundation of Shandong Province (Grants No. ZR2021QA015) and the Youth Innovation Technology Project of Higher School in Shandong Province (2023KJ146).

\end{document}